\documentclass[twocolumn,prl,superscriptaddress,floatfix,showpacs]{revtex4}
\usepackage{amssymb}
\usepackage{afterpage}
\usepackage{color}
\usepackage{epsfig}

\begin{document}

\title{Exploring structural inhomogeneities in glasses during cavitation}

\author{Pinaki Chaudhuri}
\affiliation{Institut f\"ur Theoretische Physik II, Heinrich-Heine-Universit\"at
D\"usseldorf, 40225 D\"usseldorf, Germany}
\affiliation{The Institute of Mathematical Sciences, CIT Campus, Taramani, Chennai 600 113, India}
\author {J\"urgen Horbach}
\affiliation{Institut f\"ur Theoretische Physik II, Heinrich-Heine-Universit\"at
D\"usseldorf, 40225 D\"usseldorf, Germany}


%
\begin{abstract}
Using large-scale molecular dynamics simulations for a system of $10^6$
particles, the response of a dense amorphous solid to the continuous
expansion of its volume is investigated. We find that the spatially
uniform glassy state becomes unstable via the formation of cavities,
which eventually leads to failure. By scanning through a wide range of
densities and temperatures, we determine the state points at which the
instability occurs and thereby provide estimates of the co-existence
density of the resultant glass phase.  Evidence for long-lived,
inhomogeneous configurations with a negative pressure is found, where
the frozen-in glass structure contains spherical cavities or a network
of void space.  Furthermore, we demonstrate the occurrence of hysteretic
effects when the cavitated solid is compressed to regain the dense glassy
state. As a result, a new glass state is obtained, the pressure of which
is different from the initial one due to small density inhomogeneities
that are generated by the dilation-compression cycle.
\end{abstract}

\maketitle

\section{Introduction}

A common feature of the response of amorphous solids to a mechanical
load is the formation of inhomogeneous structures that strongly affect
the materials properties of these systems \cite{schuh,greer}. Such
structures can be associated with shear banding or (micro-)cavity
formation and may lead to the initiation of crack formation or a strong
brittleness \cite{wilde,bouchaud,wang,murali,suh,falk}. Cavities
are naturally formed when a liquid is expanded such that
the two-phase region of gas-liquid coexistence is entered
\cite{balibar,joseph,oxtoby,kinjo,goddard,baidakov}.  Similarly, the
formation of cavities in dilated solids, during mechanical loading,
can be considered as a process of phase separation and in fact, for
model glass formers the metastable phase coexistence of a gas with a
glass has been envisaged \cite{tanaka} and predicted \cite{sastry}.
However, a detailed microscopic insight into the formation of cavities
in amorphous solids is lacking.

Different from the liquid-to-gas transition, one may expect that the
phase-ordering kinetics of the glass-to-gas transition is affected
by slowly relaxing (or even non-relaxing) glass domains, with a
strong dependence on the history of these domains. Thus, the usual
phase-ordering kinetics interferes with very slow relaxation processes
that are associated with structural rearrangements in the system.  On a
theoretical basis, this interplay between a slow relaxation process and
the phase separation process is only poorly understood. We note that,
on a continuum level, earlier studies have considered the latter issue
in the framework of a phenomenological model \cite{binder}.  Recent work
has explored such interplay via thermal quenches of a glass-forming
liquid into the two-phase region \cite{testard1,testard2}, investigating
the resultant morphologies.  In a silicate system, novel experimental
techniques have been recently used to visualise the spatio-temporal
evolution of such structures during similar thermal quenches leading to
phase separation \cite{bouttes,bouttes2}, with the possibility that such
studies can be extended to the glassy regimes.

In our work, molecular dynamics (MD) computer simulation are used to
investigate a model glass former that is expanded towards a metastable
miscibility gap where a glass is expected to coexist with a very
low-density gas.  During the expansion, the temperature is kept constant
such that the system always remains in a glass state until the state point
on the binodal is reached that separates the one-phase from the two-phase
region (cf.~Fig.~\ref{fig0}). Beyond the binodal, we observe the formation
of cavities, and we study their kinetics and analyze the dependence of
various thermodynamic variables as well as the structure on time.

In the context of the phase-ordering kinetics of first-order phase
transitions, different kinetic regimes can be identified that essentially
depend on the distance of the considered thermodynamic state from the
binodal of the transition. In the vicinity of the binodal, the formation
of the new phase from its mother phase is an activated process that can be
generally well described in the framework of classical nucleation theory
(CNT) \cite{reviewbinder}.  According to CNT,  the nucleation rate,
$I$, is given by $I = \kappa \exp\left[ - \frac{\Delta F^*}{k_{\rm B}T}
\right]$, with $\kappa$ a kinetic prefactor, $k_{\rm B}$ the Boltzmann
constant, $T$ the temperature, and $\Delta F^*$ the free energy barrier
for the formation of a critical nucleus. Here, the free energy barrier
$\Delta F^*$ is due to the competition between the free energy cost
for the formation of an interface between the phases on the one hand
and the free energy gain due to the formation of the new phase on the
other hand. The kinetic prefactor $\kappa$ depends essentially on the
diffusion coefficient describing the mass transport in the mother phase.
In most cases, the timescales for nucleation events, as described by
CNT, are not accessible via molecular dynamics simulations.  Further, in
the context of gas-glass phase separation,  the timescale for a cavity
to emerge would also be impacted by the extremely long relaxation
timescales of the glassy mother phase. However, if one moves farther
away from the binodal into the two-phase region, the barrier $\Delta
F^*$ decreases and eventually the description in terms of CNT is not
valid anymore. In this manner, one gradually approaches the regime
of spinodal decomposition where fluctuations leading to the formation
of the new phase spontaneously grow instead of being triggered by an
activated process. An interesting question, in this context, is how such
fluctuations and corresponding timescales are determined by the rate at
which the volume expansion is done.

The goal of our MD simulation study is to understand how a spatially
homogeneous amorphous solid in the two phase region becomes no longer
stable, if subjected to continuous volume expansion.  We observe an
abrupt formation of a cavity accompanied by a homogeneous change of
the density of the surrounding glass matrix. When keeping the total
volume of the system constant,  the kinetics of subsequent growth of the
cavity is restricted due to extremely slow structural relaxation of the
co-existing glassy material, although the system has a large negative
pressure.  As the volume expansion proceeds, the formation of cavities
eventually leads to failure and disentagration of the amorphous solids.
For each temperature, we mark the density at which cavitation occurs.
At temperatures below a threshold, related to the line of mode coupling
transition of the corresponding supercooled liquid, the cavitation density
shows little temperature dependence.  When we we compress the cavitated
solid to high densities, hysteretic effects are observed and the initial
structure of the homogeneous amorphous solid is not entirely recovered. A
new glass state is generated with small density inhomogeneities, remnant
of the cavitation process during the expansion part of the cycle.

\section{Setup and protocol}

For our study, we carry out molecular dynamics simulations of a glass
forming 80:20 binary AB Lennard-Jones (LJ) mixture, the details of which
are given in  Ref.\cite{kablj}. The intereaction potential is cutoff at
a certain interparticle distance $R_c$ and  a smoothing function is used
to ensure that both the potential and the forces continuously vanish
at the cutoff distance \cite{dyre}. All quantities are
expressed in LJ units in which the unit of length is $\sigma_{\rm AA
}$, energy is expressed in the units of $\varepsilon_{\rm AA}$, and
the unit of time is $\sqrt{{m\sigma_{\rm AA}^{2}}/\varepsilon_{\rm
AA}}$ where $m$ is the mass of the particles.
The cutoff range of the potential is chosen as $R_{c} =
2.5\sigma_{\alpha\beta}$,  with $\alpha,\beta={\rm A,B}$
referring to the type of particles considered as a pair.

\begin{figure}
\includegraphics[scale=0.45,angle=0,clip=true]{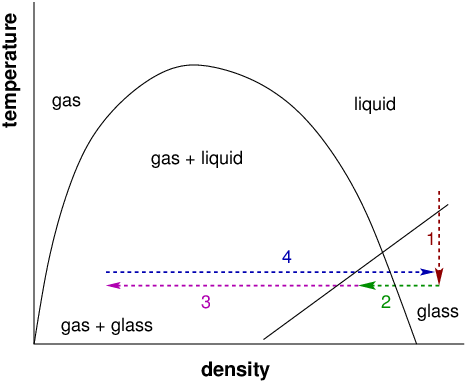}
\caption{\label{fig0} Schematic diagram indicating the various paths
traversed across the $T-\rho$ phase diagram in our numerical experiments.
(1) - quench from supercooled liquid; (2) - expansion of amorphous
solid until cavitation is observed; (3) - expansion of cavitated solid
to explore low density states; (4) - compression of the cavitated solid
back to initial high density.}
\end{figure}

We prepare the amorphous solid by instantaneous quenches from the
supercooled liquid. Most of our results are reported for a density of
$\rho=1.3$ where mechanically stable states, i.e.~with positive bulk
pressure, exist to very low temperatures. Quenches to target temperatures
($T=0.5$, 0.4, 0.3, 0.2, 0.1, 0.01) are done from equilibrated supercooled
liquid states ($T=0.66$) \cite{ludo}. For some comparative studies,
we also do simulations at $\rho=1.2$ and $\rho=1.4$. In those cases, the
instantaneous quenches are done from $T=0.44$ and $T=0.945$, respectively,
where the equilibrium diffusion constants are approximately similar to
that measured at $T=0.66$ for $\rho=1.3$.  Such quenches correspond to
the path 1 marked in the schematic diagram shown in Fig.~\ref{fig0}.

After the quench to the target temperature, the glass is aged for $t_{\rm
w}=10^4$, before we begin our expansion experiments. Once aged, we carry
out the following protocol for volume expansion.  At every expansion step,
we scale up the linear size of the cubic simulation box by a factor
of $\Delta{L}=1.001$ to obtain smaller densities. With the particle
co-ordinates also getting similarly rescaled, this corresponds to an
affine transformation of the system.  Subsequently, we allow the system
to relax over a time period of $\delta{t}=10^3$, before the next expansion
step is carried out. Some expansions are also done for smaller $\delta{t}$
in order to compare the effects of the expansion rate. These expansion
steps, at fixed temperatures, follow the paths 2 and 3, as marked on
the schematic diagram (Fig.~\ref{fig0}). We also report results from
compression experiments, where the the linear size of the box is scaled
by a factor of $\Delta{L}=0.9990$, with intermediate $\delta{t}=25$.
This follows path 4 in Fig.~\ref{fig0}, with the objective of returning
back to the initial density from where the expansion of the solid was
initially started.  The expansion (or compression) protocol, described
above, would correspond to ramp experiments. This is thus distinct from
the protocol used in Ref.~\cite{falk}, where the solid was first expanded
to certain strain amplitudes and thereafter the timescales for a cavity
to emerge were obtained.

\begin{figure*}[t]
\includegraphics[scale=0.55,angle=0,clip=true]{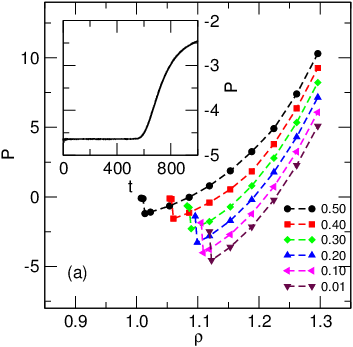}
\includegraphics[scale=0.38,angle=0,clip=true]{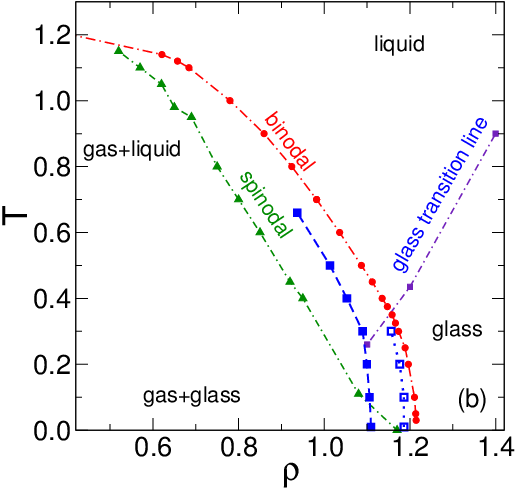}\\
\includegraphics[scale=0.55,angle=0,clip=true]{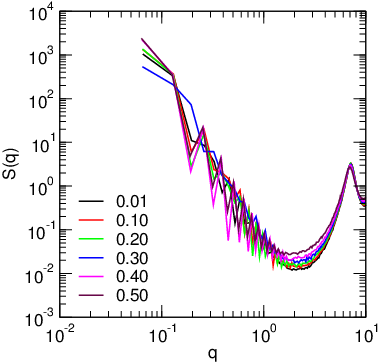}
\includegraphics[scale=0.55,angle=0,clip=true]{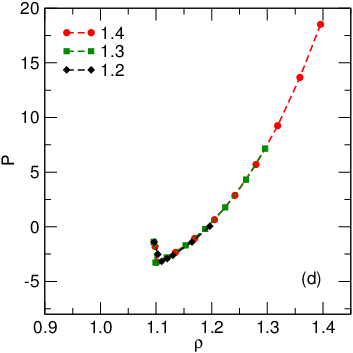}
\caption{\label{fig1} (a) Evolution of pressure ($P$) with decreasing
density $\rho$ for the different temperatures $T=0.01$, 0.1, 0.2, 0.3,
0.4, 0.5, with expansions starting from $\rho_0=1.3$. The inset shows the
rapid increase in pressure when cavitation occurs inside the material;
data is shown for $T=0.01$. (b) For each temperature $T$, the density
at which cavitation occurs is marked on the phase diagram by filled blue
squares (the data is obtained from Ref.~\cite{testard2}).  We also mark,
using empty blue squares, the density of the amorphous solid which is
in co-existence with the gas phase.  (c) For different temperatures, the
partial structure factor $S_{\rm AA}(q)$ after cavitation is displayed.
(d) $P$ vs.~$\rho$ at $T=0.2$ for states having initial densities
$\rho_0=1.2$, 1.3, 1.4.}
\end{figure*}
Thermostatting at any temperature ($T$) is done by sampling velocities,
at every 400 time steps, from the corresponding Maxwell-Boltzmann
distribution. Here, we note that our expansion process is done under
thermostatted conditions, unlike the micro-canonical $NVE$ ensemble used
in Ref.~\cite{falk} (with $E$ the total energy).  Due to thermostatting,
there is no runaway heating as is the case in the $NVE$ ensemble, due
to the conservation of the total energy $E$.

The MD simulations are done for a system size of $N=10^6$ particles, using
the GPU-accelerated code HALMD \cite{felix}. For each state point that
we have analysed, we do averages over $m=10$ independent trajectories.

\section{Results}
\subsection{Onset of cavity}
To monitor the thermodynamic state of the system during the expansion
process (path 2 in Fig.~\ref{fig0}), we measure the bulk pressure. With
every step of expansion, the pressure drops as the particles are
pushed apart.  After each instantaneous expansion step, the pressure
stabilises during the subsequent waiting time $\delta{t}$.  As we decrease
the density step by step, the pressure becomes negative implying the
presence of internal tension. At some point during the expansion, the
system can no longer sustain the built-up tension and the  homogeneous
system becomes unstable, with the occurrence of  cavitation via a sudden
increase in pressure [inset of Fig.~\ref{fig1}(a)], similar to what
has been observed in liquids \cite{goddard}. This happens for all the
configurations for the range of temperatures that we have studied. The
variation of pressure with decreasing density and the onset of cavitation
are shown in Fig.~\ref{fig1}(a).  We note here that the shape of the $P$
vs.~$\rho$ isotherms are similar to those obtained by Sastry \cite{sastry}
for supercooled liquid states.

We record the density at which cavitation occurs for each of the $m$ runs
at each temperature $T$. The ensemble-averaged density for the onset of
cavitation is marked on the phase diagram obtained by Testard {\it et
al.}~\cite{testard1, testard2}. We note that the {\it cavitation line}
[see Fig.~\ref{fig1}(b)] obtained by us lies to the left of the binodal
line obtained in this earlier work.  In the case of Testard {\it et
al.}~\cite{testard1}, the binodal line was obtained by measuring the
density of the liquid/glass phase obtained during phase separation via
thermal quenches. For us, the {\it cavitation line} marks the occurrence
of the cavity within the time window of $\delta{t}=10^3$ between each
expansion step.  We also further observe that, at low temperatures,
i.e.~below the extrapolated MCT-line \cite{ludo}, the density at which
cavitation occurs does not differ much with $T$. However, as this MCT-line
is crossed, thermal effects allow for the relaxation of local stresses
built up during the expansion process, thus increasingly delaying the
emergence of cavities.

At the density at which cavitation occurs, we measure the partial
structure factor corresponding to the correlations between particles of
type $\textrm{A}$ of the binary mixture, $S_{\rm AA}(q)$, at the end
of the time window $\delta{t}$ (for a definition of $S_{\rm AA}(q)$,
see Ref.~\cite{glassbook}). For all temperatures, we observe a similar
low-$q$ divergence in $S_{\rm AA}(q)$ [see Fig.~\ref{fig1}(c)], indicating
the formation of a phase separated state \cite{bray}.

\begin{figure*}
\includegraphics[width=7cm,angle=0,clip=true]{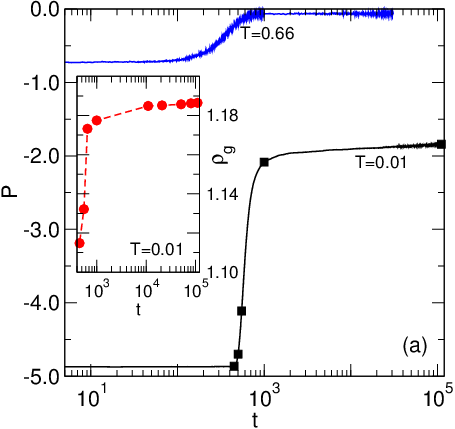}
\includegraphics[width=7.2cm,angle=0,clip=true]{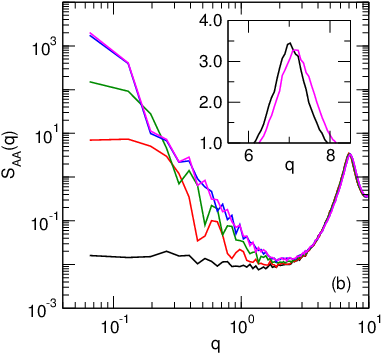}
\includegraphics[width=14cm,angle=0,clip=true]{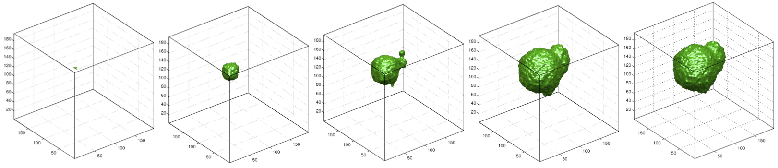}
\caption{\label{fig2} ({\bf Top}) (a) At $T=0.01$, (in black) evolution
of pressure with time, after cavitation, if the expansion is stopped
(i.e.~global density remains fixed). Also shown is the corresponding
data for $T=0.66$ (in blue). In the inset, the density of the amorphous
solid co-existing with the low density phase inside the cavity is shown
for $T=0.01$.  (b) Evolution of structure factor: $S_{\rm AA}(q)$
measured at different times as marked in the main panel of (a) for
$T=0.01$.  The inset shows that there is a shift in the location of
the first sharp diffraction peak of $S_{\rm AA}(q)$ between $t=450$
and $t=110000$, indicating a densification.  ({\bf Bottom}) Growth of
cavity as a function of time: density plot showing evolution of the
gas-solid interface at the above marked times.}
\end{figure*}
We also check whether the density of cavitation depends upon the initial
density of the quenched amorphous solid, from where the expansion process
is started. Thus, at $T=0.2$, we start the expansion from two other
densities, viz.~$\rho_0=1.2$ and $\rho_0=1.4$. Note that, for $\rho=1.2$, at
temperatures lower than $T=0.2$, quenched amorphous states are unstable,
i.e.~they have negative pressure. Therefore, at this density, our dilation
experiments are done for $T=0.2$, such that we have stable homogeneous
structures as initial states.  We show in Fig.~\ref{fig1}(d) that the
onset of cavitation does not depend on $\rho_0$. We have checked this to
be the case for the ensemble of $m$ initial states. Thus, the point of
instability of the homogeneous amorphous solid does not seem to depend
upon the amount of deformation undergone via the expansion protocol.

\subsection{Density of co-existing glass}
For properly marking out the boundaries of the phase diagram, knowing the
density of the co-existing phases is required.  Thus, we try to estimate
the density of the glass ($\rho_{\rm g}$) which is in co-existence
with the low-density vapour phase.  In order to measure that, we stop
the expansion of the solid as soon as we identify that a cavity has
formed in the system. We then proceed with the usual particle dynamics,
holding the density constant at the cavitation density, and observe how
the structure evolves with time.

In Fig.~\ref{fig2}(a), for the glassy state at $T=0.01$, we see that
after the initial rapid increase of pressure, the rate of increase
slows down and nearly saturates at long times. The evolution of the
structure is also minimal, as can be seen from the $S_{\rm AA}(q)$ at
early and later times [see Fig.~\ref{fig2}(b)]. The power-law divergence
has an exponent of $3.74$ which is close to Porod's law (corresponding
to a smooth interface) \cite{bray}. To further illustrate the evolution
of the structure, we compute the density iso-surfaces corresponding to
$\rho=0.2$ at several values of $t$ [as marked in Fig.~\ref{fig2}(a)] --
this iso-surface encloses the low-density cavity. We see the emergence of
the cavity as the jump in pressure initiates. When the pressure rapidly
increases to release the local stresses, the cavity expands rapidly. At
later times, the growth of the cavity slows down, when the time evolution
of the pressure becomes very slow. We note that the pressure is still
negative. While the system would like to release the residual tension
to reach the non-negative co-existence pressure, the glassy dynamics
of the surrounding makes this process extremely slow. For comparison,
we also show in Fig.~\ref{fig2}(a) the time evolution of pressure during
cavitation in a liquid state at $T=0.66$. Firstly, the pressure values
are much higher and unlike $T=0.01$, the jump in pressure during the
formation of cavity is smaller and also slower. Also, asymptotically,
for the case of the liquid, the system is closer to a positive pressure
state, unlike the glassy system.

As the cavity emerges, we start to measure the density of the amorphous
solid which co-exists with the cavity.  At every time of interest
[marked in Fig.~\ref{fig2}(a)], the cubic box is divided into smaller
boxes having a linear dimension of $10$ diameters. The local density
is calculated in each small cube and the distribution is plotted. When
phase separation occurs, we get a bimodal distribution; $\rho_{\rm g}$
corresponds to the mean value of the distribution at high density. In
the inset of Fig.~\ref{fig2}(a), we show how $\rho_{\rm g}$ evolves with
the time -- the glassy matrix surrounding the cavity suddenly densifies
following the emergence of a void. This densification process is also
captured in $S_{\rm AA}(q)$, as shown in the inset of Fig.~\ref{fig2}(b);
we see that, after cavitation, the location of the peak shifts to a
larger wave-vector, indicating the emergence of a more dense solid.

\begin{figure*}
\includegraphics[width=7cm,angle=0,clip=true]{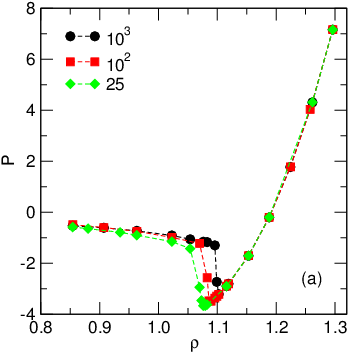}
\includegraphics[width=7.2cm,angle=0,clip=true]{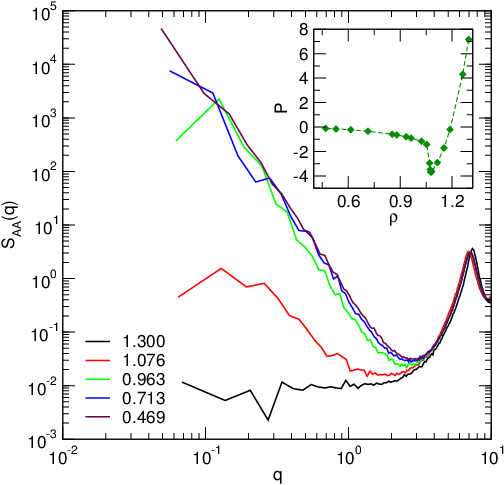}\\
\includegraphics[scale=0.7,angle=0,clip=true]{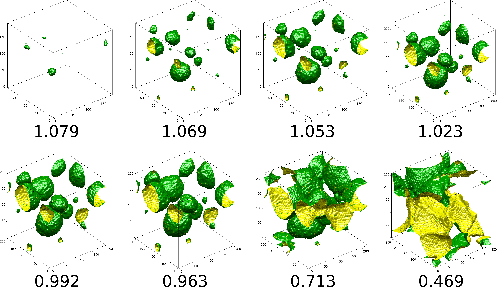}\\
\caption{\label{fig3} {\it Continuing expansion}: ({\bf Top}) $T=0.2$:
(left) evolution of pressure for different $\delta{t}=1000$ (black),
100 (red), 25 (green). (Right) $S_{\rm AA}(q)$ for $\rho=1.3$, 1.076,
0.963, 0.713, 0.469. The inset shows the $P$ vs.~$\rho$ data for the
corresponding path across the phase diagram. ({\bf Bottom}) Evolution
of density field for $\delta{t}=25$, with the green and yellow colours
representing respectively the solid and gas sides of the interfaces.}
\end{figure*}
For temperatures below the MCT line, we mark this asymptotic value
of $\rho_{\rm g}$ in the phase diagram [Fig.~\ref{fig1}(b) (using
empty squares)].  The dependence on temperature is similar to our {\it
cavitation line}. In accordance with Ref.~\cite{testard2}, we also find
that there is no low temperature re-entrant behaviour in $\rho_{\rm
g}$ as suggested by Cardinaux {\it et al.}~\cite{cardinaux}. We also
note that our estimates of $\rho_{\rm g}$ are slightly smaller than
those estimated in Ref.~\cite{testard2} -- in our case, the existence
of one or two cavities allows for easier estimation than compared to
estimations from more complex structures as is the case for Testard {\it
et al.}~\cite{testard2}.

\subsection{Exploring lower density states via expansion}
In order to explore the process of large-scale failures in amorphous
solids during dilation, we continue the expansion process beyond
densities where cavities form (path 3 in Fig.~\ref{fig0}). We wish
to investigate how larger and more complex void spaces emerge as
we continue the dilation.  In order to expedite the exploration of
a large range of densities within the simulational time scales, we
resort to using a smaller $\delta{t}$, i.e.~a smaller waiting time
after each expansion step. Thus, we first check the effect of changing
$\delta{t}$. As demonstrated in Fig.~\ref{fig3}(a) for $T=0.2$, the onset
of cavitation is slightly delayed, i.e.~with decreasing $\delta{t}$,
it happens at lower densities $\rho$.  Further, the jump in pressure at
cavity formation is also rounded off. A smaller $\delta{t}$ allows for
less time for the structure to adjust to the expansion process, which
delays the eventual failure. This is consistent with the expectation that
the time scale for cavitation becomes increasingly smaller as we move
into the phase-coexistence region away from the binodal line. However,
for all $\delta{t}$, once cavitation has occurred in each case, the
pressure again increases with decreasing density as the system eventually
tries to attain a stable state (corresponding to a positive pressure).
Eventually, the $P$ vs.~$\rho$ curves become nearly identical for the
different $\delta{t}$ we have explored, implying that the eventual stable
states at smaller $\rho$ do not depend on the rate of expansion.

\begin{figure*}
\begin{center}
\includegraphics[width=15.5cm,angle=0,clip=true]{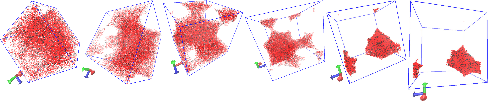}
\end{center}
\caption{\label{fig4} Sequence of snapshots during volume expansion at
$T=0.2$. From left: $\rho=0.963$, 0.714, 0.529, 0.392, 0.113, 0.057.}
\end{figure*}
For $\delta{t}=25$, we travel to lower densities in the gas-glass
coexistence region.  We note that at densities around and less than
$\rho=0.469$, the pressure is nearly zero.  In Fig.~\ref{fig3}(b),
we show the evolution of the structure, by measuring $S_{\rm AA}(q)$,
with decreasing $\rho$. To further visualise the evolution of the
structure, we show the shape of the density field corresponding to the
low-density phase that is in co-existence with the amorphous solid
(see bottom panel of Fig.~\ref{fig3}). We use the following colour
code - the yellow iso-surface, corresponding to $\rho=0.1$, faces the
gas and the green iso-surface, corresponding to $\rho=0.2$, faces the
solid. Thus, initially, at high densities we have nearly spherical
cavities emerging (cf.~the snapshot for $\rho=1.069$).  As the density
is decreased, cavities merge and take more and more non-spherical shapes
(cf.~the snapshot for $\rho=0.963$). Eventually, via the merger of these
enclosed spaces, a interconnected structure emerges at lower densities
(cf.~the snapshot for $\rho=0.469$). We note here that in experiments,
the spatial percolation of the void space would correspond to the complete
disintegration of the amorphous solid.

We have expanded the system further in order to estimate the density
where the co-existence line is crossed in the low-density limit
and a homogenous phase is observed, i.e.~continuing along path 3 in
Fig.~\ref{fig0}.  The sequence of states is illustrated using snapshots
of particle configurations; see Fig.~\ref{fig4}. As we explore smaller
and smaller densities, eventually we obtain an amorphous cluster
which is in co-existence with the gas (see, e.g., the configuration at
$\rho=0.113$). We note that the cluster is not spherical in shape --
one would expect that the forces due to surface tension would lead to
smoothening out the surface of the cluster.  However, because we are at
low temperatures, the structural relaxation time scales are enormously
large and therefore the interface does not relax further, leading to
faceted surfaces. We have checked and found that even if we stop the
expansion process and wait for long time scales, the shape of the cluster
does not change due to the slow relaxation time scales. This is unlike
the case for the liquid state where eventually a liquid droplet emerges
to minimise the interfacial energy \cite{reviewbinder}.  Finally, we
note that even at densities as low as $\rho=0.057$, the system remains
phase-separated, i.e.~the homogeneous phase corresponds to even lower
densities, the exploration of which is beyond our current simulational
possibilities. This is consistent with the phase diagram obtained by
Testard {\it et al.}~\cite{testard2}, where we notice that the low
temperature regime in the low density binodal is missing, possibly due
to similar numerical difficulties.

\subsection{Compressing the cavitated solid}
\begin{figure*}
\includegraphics[width=14cm,angle=0,clip=true]{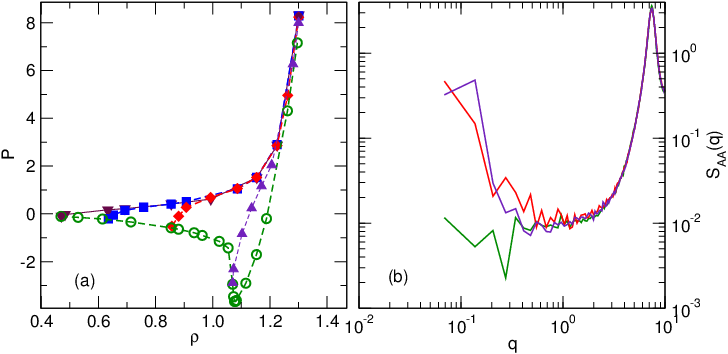}\\
\includegraphics[width=14cm,angle=0,clip=true]{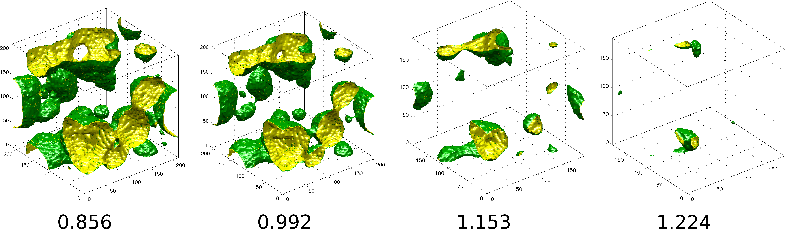}\
\caption{\label{fig5} ({\bf Top}) (a) For $T=0.2$ and $\delta{t}=25$,
evolution of the pressure with density (filled symbols) during
compression, starting from different low densities (different colors). The
corresponding $P$-$\rho$ data for the expansion path is shown as empty
symbols. (b) $S_{\rm AA}(q)$ at $\rho=1.30$ for the quenched amorphous
state (in green) and the solid formed via compression starting from
$\rho=0.856$ (in red) and $\rho=1.069$ (in purple). ({\bf Bottom})
Density plots demonstrating how void spaces decrease in size as the
material is compressed.}
\end{figure*}

Finally, we inquire how the material responds if the cavitated solid
is compressed back towards the high density from where expansion
was initiated.  Starting from different low density states generated
during the expansion, we carry out this compression process at fixed
temperature (path 4 in Fig.~\ref{fig0}). During the compression,
we choose $\delta{t}=25$. Data for the corresponding evolution of
pressure with density is shown in Fig.~\ref{fig5}(a) (using filled
symbols). When the material is compressed, the pressure continually
increases, as expected. We see that, apart from some initial regime,
all these different initial low density states follow the same monotonic
isotherm when the solid is compressed. Thus, if one considers the full
expansion-compression cycle, there is a hysteretic effect which occurs
via the cavitation process -- the longer we wait before we begin the
compression, the larger is the area under the loop.

In Fig.~\ref{fig5} (bottom), we visualise the low density regions
inside the system, for the case where we compress the material from
$\rho=0.856$. With increasing density, the voids slowly shrink and
eventually when the binodal is crossed the voids nearly disappear;
see the density field for $\rho=1.224$ in Fig.~\ref{fig5} (bottom).
Going back to the isotherms shown in Fig.~\ref{fig5}(a), we see that
the variation of pressure with density becomes steeper beyond these
densities. Once we reach the initial density ($\sim \rho=1.3$) from
where the expansion was started, we notice that the pressure is slightly
higher than that of the undeformed quenched state. In fact, the partial
structure factor [Fig.~\ref{fig5}(b)] shows that the low-$q$ behavior
is different, implying that vestiges of the cavitation process (in the
form of density inhomogeneities) have remained within the structure. The
pressure thus increases to accommodate this structural change. Comparing
two cases of compression beginning from $\rho=1.069$ and $\rho=0.856$,
we find that the degree of mismatch in structures is more pronounced
for the lower initial density.

\section{Conclusions}
We have studied the process of phase separation in a model amorphous
solid, via the continuous ramping up of its volume. We observed that,
upon crossing the binodal line in the phase diagram, the spatially uniform
glass becomes unstable, leading to the formation of cavities. For a given
temperature, the density for cavitation depends upon the rate at which
the expansion is done. This implies that the onset of phase separation
is due to an interplay between the distance from the binodal and the time
scale allowed for the structure to relax the increasing internal tension
during successive dilation steps. For temperatures below the mode coupling
transition line, thermal effects seem insignificant in determining the
density for cavitation to occur, at a fixed expansion rate. However, above
the line, thermal fluctuations allow faster relaxation time scales, thus
delaying the onset of phase separation to increasingly smaller densities.

Similar to what has been reported in the context of crystal-to-glass
transitions \cite{sciortino95}, we find the possibility of long-lived
inhomogeneous glassy structures at negative pressures.  This is different
from gelation in systems with short-ranged attraction where the formation
of a gel suppresses the onset of phase separation \cite{lu08}. In the
case of cavity formation in the glass, the extremely long relaxation
time scales of the glassy matrix surrounding the cavities or the network
of voids provide the long life-time of inhomogeneous structures. Such
glassy time scales also lead to the formation of non-spherical amorphous
clusters at very low densities, when the solid is further expanded. At
some intermediate density, the void space percolates through the system,
which would correspond to the failure of the solid in a real experiment.

We also explored the possibility of regaining the glassy state, if
we compress the solid with cavities.  We find that a single cycle of
expansion and compression leads to hysteretic effects. There is a mismatch
of structures at the initial density from where the cycle originated --
the memory of the cavitation process is retained via the formation of
a new inhomogeneous glassy state, with a higher pressure, at the end of
the cycle.  As intermediate states during compression, we obtain porous
structures which are mechanically stable. Thus, such expansion-compression
paths are also possible routes to form amorphous porous solids.

Finally, we would like to note that one has to be careful when
interpreting the onset of phase separation in the framework of classical
nucleation theory.  As we discussed earlier, the dynamic regime that is
accessible by MD simulations is usually associated with the early stages
of spinodal decomposition, in our case accompanied by the slow relaxation
of the glass surrounding the cavity.  A description in terms of CNT is,
however, only sensible if the free energy barrier for the formation of
a critical nucleus is much larger than the thermal energy, i.e.~$\Delta
F^* >> k_{\rm B}T$.  In the work of Guan {\it et al.}~\cite{falk}, the
measured free energy barriers for the cavity to form in a metallic glass
are at most of the order of $10 k_{\rm B}T$.  For such small barriers,
cavity formation cannot be treated as an activated process, in contrast
to the analysis by Guan {\it et al.} Further analysis is needed in
understanding the cavitation in glasses, specially how barriers of
phase separation interplay with the external stresses applied during
mechanical loading.

\section{Acknowledgements}
We thank the DFG research unit SPP 1594 (project HO 2231/8-1) for
financial support. PC thanks the Starting Ramp Program within SPP 1594
for funding of GPU hardware.  We also thank L. Berthier, M. Falk, W. Kob,
and S. Sastry for useful discussions and N. H\"oft for the numerical
analysis using HALMD.


\begin{thebibliography}{99}
%
\bibitem{schuh}
C. A. Schuh, T. C. Hufnagel, and U. Ramamurty, 
Acta Mater. {\bf 55}, 4067 (2007).
%
\bibitem{greer}
M. F. Ashby and A. L. Greer,
Scripta Mater. {\bf 54}, 321 (2006).
%
\bibitem{wilde}
G. Wilde, J. H. Chen, C. B. Qu, S. Y. Fu, and F. Jiang,
Acta Mater. {\bf 77}, 248 (2014).
%
\bibitem{bouchaud}
E. Bouchaud, D. Boivin, J. L. Pouchou, D. Bonamy, B. Poon, and G. Ravichandran,
EPL {\bf 83}, 66006 (2008).
%
\bibitem{wang}
G. Wang, D. Q. Zhao, H. Y. Bai, M. X. Pan, A. L. Xia, B. S. Han, 
X. K. Xi, Y. Wu, and W. H. Wang,
Phys. Rev. Lett. {\bf 98}, 235501 (2007).
%
\bibitem{murali}
P. Murali, T. F. Guo, Y. W. Zhang, R. Narasimhan, Y. Li, and H. J. Gao,
Phys. Rev. Lett. {\bf 107}, 215501 (2011).
%
\bibitem{suh}
J.-Y. Suh, R. D. Conner, C. P. Kim, M. D. Demetriou, and W. L. Johnson, 
J. Mater. Res. {\bf 25}, 982 (2010).
%
\bibitem{falk}
P. Guan, S. Lu, M. J. B. Spector, P. K. Valavala, and M. L. Falk,
Phys. Rev. Lett. {\bf 110}, 185502 (2013).
%
\bibitem{balibar}
H. Maris and S. Balibar,
Physics Today {\bf 53}, 29 (2007).
%
\bibitem{joseph}
D. D. Joseph, 
J. Fluid Mech. {\bf 366}, 367 (1998).
%
\bibitem{oxtoby}
D. W. Oxtoby, Acc. Chem. Res. {\bf 31}, 91 (1998).
%
\bibitem{kinjo}
T. Kinjo and M. Matsumoto,
Fluid Phase Equil. {\bf 144} 343 (1998).
%
\bibitem{goddard}
Q. An, G. Garrett, K. Samwer, Y. Liu, S. V. Zybin, S.-N. Luo, M. D. Demetriou, 
W. L. Johnson, and W. A. Goddard III, 
J. Phys. Chem. Lett. {\bf 2}, 1320 (2011).
%
\bibitem{baidakov}
V. G. Baidakov and K. S. Bobrov,
J. Chem. Phys. {\bf 140}, 184506 (2014).
%
\bibitem{tanaka}
H. Tanaka, 
J. Phys.: Condens. Matter {\bf 12}, R207 (2000).
%
\bibitem{sastry}
S. Sastry,
Phys. Rev. Lett. {\bf 85}, 590 (2000).
%
\bibitem{binder}
K. Binder, H. L. Frisch, and J. J\"ackle,
J. Chem. Phys. {\bf 85}, 1505 (1986).
%
\bibitem{testard1} 
V. Testard, L. Berthier, and W Kob,
Phys. Rev. Lett. {\bf 106}, 125702 (2011).
%
\bibitem{testard2}  
V. Testard, L. Berthier, and W. Kob,
J. Chem. Phys. {\bf 140}, 164502 (2014).
%
\bibitem{bouttes}
D. Bouttes, E. Gouillart, E. Boller, D. Dalmas, and D. Vandembroucq,
Phys. Rev. Lett. {\bf 112}, 245701 (2014).
%
\bibitem{bouttes2}
D. Bouttes, O. Lambert, C. Claireaux, W. Woelffel, D. Dalmas, E. Gouillart,  
P. Lhuissier, L. Salvo, E. Boller, and D. Vandembroucq,
Acta Mater. {\bf 92} 233 (2015).
%
\bibitem{reviewbinder}
K. Binder, 
Rep. Prog. Phys. {\bf 50} 783 (1987).
%
\bibitem{kablj}
W. Kob and H. C. Andersen,
Phys. Rev. E {\bf 51}, 4626 (1995).
%
\bibitem{dyre}
S. Toxvaerd, O. J. Heilmann, and J. C. Dyre,
J. Chem. Phys. {\bf 136}, 224106 (2012).

%
\bibitem{felix}
P. H. Colberg and F. H\"ofling,
Comp. Phys. Comm. {\bf 182}, 1120 (2011).
%
\bibitem{ludo}
L. Berthier and G. Tarjus
Phys. Rev. E {\bf 82}, 031502 (2010).
%
\bibitem{glassbook}
K. Binder and W. Kob,
{\it Glassy Materials and Disordered Solids: An Introduction to Their
Statistical Mechanics}
(World Scientific, Singapore, 2011).
%
\bibitem{bray}
A. J. Bray, 
Adv. Phys. {\bf 43}, 357 (1994). 
%
\bibitem{cardinaux}
F. Cardinaux, T. Gibaud, A. Stradner, and P. Schurtenberger, 
Phys. Rev. Lett. {\bf 99}, 118301 (2007).
%
\bibitem{sciortino95}
F. Sciortino, U. Essmann, H. E. Stanley, M. Hemmati, J. Shao, 
G. H. Wolf, and C. A. Angell,
Phys. Rev. E {\bf 52}, 6484 (1995). 
%
\bibitem{lu08}
P. J. Lu, E. Zaccarelli, F. Ciulla, A. B. Schofield, F. Sciortino, 
and D. Weitz,
Nature {\bf 453}, 499 (2008). 
\end{thebibliography}
\end{document}